\documentclass[11pt]{iopart}

\usepackage{iopams}
\usepackage[latin1]{inputenc}
\usepackage{amsfonts}
\usepackage{amssymb}
\usepackage{graphicx}
\usepackage{cite}
\usepackage{color}
\usepackage{bbold}
\usepackage{enumerate}
\usepackage{hyperref}
\usepackage{libertine}

\newcommand{\av}[1]{\langle{#1}\rangle_{\rm s}}

\newcommand{\sym}[2]{\chi^{#1}_{#2}}

\newcommand{\gen}{{\bf M}}
\newcommand{\pp}{{\bf p}}

\begin{document}
\title[Phase transitions and departure statistics of critically loaded queues]
{\large Phase transitions and departure statistics of critically loaded queues:
oscillating cumulants and generalized BRAVO}

\author{M.~Bruderer}
\address{Wenger Engineering GmbH, Research and Development, Einsteinstraße~55, 89077 Ulm, Germany}
\ead{martin.bruderer@wenger-engineering.com}

\date{\today}

\begin{abstract}
Queueing theory is used for modeling biological processes, traffic flows and many more
real-life situations. Beyond that, queues describe systems out of equilibrium and can thus be considered
as minimal models of non-equilibrium statistical mechanics. We demonstrate that non-equilibrium
phase transitions of queues in the steady state are accompanied by a nontrivial flow of
departing customers. Our analytical results show that the cumulants of the departure statistics
deviate strongly from Poissonian values and oscillate in the vicinity of phase transitions,
i.e., if a critical load is approached. The load-dependent oscillations of the cumulants
generalize the BRAVO effect (Balancing Reduces Asymptotic Variance of Outputs) in queues and
may occur in other boundary-driven non-equilibrium systems.
\end{abstract}

\section{Introduction}

A central topic of non-equilibrium statistical mechanics are driven transport
models~\cite{bodineau2007cumulants,derrida2007non,mallick2015exclusion}. These models
show a surprisingly complex behavior and exhibit interesting effects such as
non-equilibrium phase transitions and intricate statistics of the current of particles.
There are only few examples of non-equilibrium models for which the statistics in terms
of cumulants or the large deviation function of the current can be calculated. However,
some models are tractable by analytical methods and have been analyzed in detail.
A paradigmatic example is the totally asymmetric exclusion process
(TASEP)~\cite{Derrida-PRL-1998,Derrida2004,Evans-PA-2002,Golinelli-JPA-2006,Chou-RPP-2011}.
The TASEP describes boundary-driven particle hopping on a lattice, where particles cannot
occupy the same lattice site. Closely related to the TASEP is the exclusive queueing process
(EQP)~\cite{arita2011dynamical,arita2014dynamics}, which in contrast to
standard queueing models takes the excluded-volume effect into account. In fact, the stationary
state of a parallel-update TASEP with varying system length and the EQP are equivalent. For both
models the exact solutions are known for the stationary state and their phase diagrams have been
extensively studied.

\begin{figure}[h]
\centering
\raisebox{25pt}{\includegraphics[width=150pt]{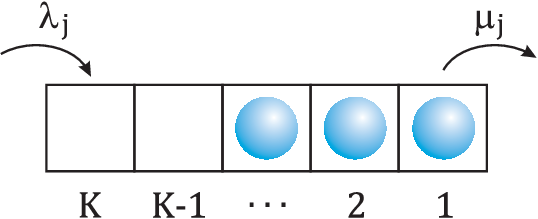}}\hspace{30pt}\includegraphics[width=90pt]{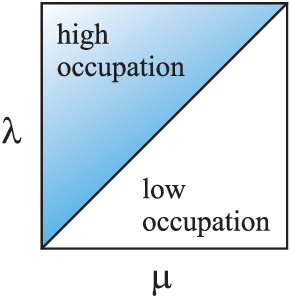}
\caption{Customers enter and leave a queue with rates $\lambda_j$ and $\mu_j$, respectively. The waiting space of the queue
is of finite size $K$ so that the arrival of customers is blocked if the waiting space is full. The occupation of the waiting
space allows us to identify a non-equilibrium phase transition between a high- and low-occupation phase of the $M/M/1/K$ queue,
depending on the rates
$\lambda$ and $\mu$.}
\label{scheme}
\end{figure}

We show in this paper that even standard queueing models display several interesting non-equilibrium features by
combining ideas of non-equilibrium statistical mechanics and queueing theory~\cite{Kleinrock-1976,van2014performance}.
While conventional queueing models are usually applied to practical problems~\cite{leibowitz1968queues,van2007modeling,arita2015exclusive,romano2009queueing,hochendoner2014queueing}
they are sufficiently versatile to be of physical relevance and often permit exact analytical solutions.
The basic quantities of interest in non-equilibrium statistical mechanics and queueing theory
are the same: The density of particles corresponds to the occupation of the queue, and the
counting statistics of the current through the system translates into the statistics of the
queue departure process.

We consider single server queues, where customers enter and leave the waiting space of size $K$ with
time-independent arrival and departure rates (cf.~Fig.~\ref{scheme}). If the waiting space is full then
the arrival of customers is blocked and customers are lost. The specific service policy for the queueing
model (e.g.~first in, first out) is not relevant in this context. The system exhibits a non-equilibrium
phase transition between phases of high and low occupation of the waiting space in the limit of large $K$,
which is induced by the driving rates and accompanied by enhanced fluctuations of the length of the queue.
We show that this non-equilibrium phase transitions are directly reflected in the counting statistics of
the flow of customers through the queueing system.
A fundamental result in this context is Burke's theorem~\cite{Burke-OR-1956,Burke-AMS-1968}
stating that the steady-state departure process of a queue with infinite waiting space obeys
Poissonian statistics. We will see that the departure statistics for finite waiting spaces can
deviate markedly from Poisson statistics close to phase transitions.

We find that the cumulants of the counting statistics oscillate strongly in the vicinity of
the phase transitions and reach a local minimum or maximum exactly at the transition point.
We expect that such oscillations of the cumulants of the departure process are also observable
close to phase transitions of other queueing models and, more generally, in other
boundary-driven non-equilibrium systems.

\section{Characterizing queueing models}\label{general}

Let us denote by $N(t)\in\{0,\ldots,K\}$ the number of customers in the queue at time $t\geq 0$, and similarly
$n(t)\equiv N(t)/K$ is the occupation of the waiting space. We further introduce  the probabilities $p(j,t)$
of finding $N(t) = j$ at time $t$ and the corresponding column vector $\pp(t) = \{p(j,t),\, j=0,\ldots,K\}$.
The arrival and service of customers are described by a Poisson process with rate $\lambda_j$ and $\mu_j$,
respectively, where the index $j\in\{0,\ldots,K\}$ allows for a dependence on $N(t)$. The time evolution
of the queue is governed by the continuous-time Markovian master equation $\rmd\pp(t)/\rmd t = \gen\pp(t)$
with the time-independent generator 
\begin{equation}\label{generator}
\gen = \left(
  \begin{array}{ccccc}
    -\lambda_0 & \mu_1 &  &  & 0 \\
    \lambda_0 & -(\lambda_1 + \mu_1) & \mu_2  &  & \\
     & \ddots & \ddots & \ddots & \\
     &  & \lambda_{K-2} & -(\lambda_{K-1} + \mu_{K-1}) & \mu_{K} \\
     0 &  &  & \lambda_{K-1} & -\mu_{K}
  \end{array}
\right),
\end{equation}
which has a birth-death structure on the state space $\{0,\ldots,K\}$.

\subsection{Non-equilibrium steady-state occupations and flows}

In the following we focus on the stationary evolution of the system and assume that it has a unique
non-equilibrium steady state (NESS). This state is specified by the stationary probability distribution $\pp_{\rm s}$
defined by $\gen\pp_{\rm s} = 0$, i.e., the generator $\gen$ has a unique zero eigenvalue $\omega_0=0$. To
identify different phases of the queueing system we choose a suitable order parameter, namely the average occupation
in the steady state $\av{n}$, which takes values in the interval $[0,1]$. Additional information is provided by the
variance $\av{\delta n^2} = \av{n^2} - \av{n}^2$ and higher-order moments. These quantities are found from the
steady-state solution of the master equation~\cite{Kleinrock-1976}
\begin{equation}\label{gensol}
\eqalign{
	p_{\rm s}(0) &= \left(1 + \sum_{\ell=1}^K\prod_{i=0}^{\ell-1}\frac{\lambda_i}{\mu_{i+1}}\right)^{-1} \\
	p_{\rm s}(j) &= p_{\rm s}(0)\prod_{i=0}^{j-1}\frac{\lambda_i}{\mu_{i+1}}\qquad j\geq 1
}
\end{equation}
together with the definition of the moments $\av{n^\ell} = \sum_{j}^K (j/K)^\ell p_{\rm s}(j)$. In some cases
it is possible to evaluate the sums explicitly to find closed-form expressions for the lowest moments $\av{n^\ell}$.


For characterizing the steady-state flow of customers through the queue we consider the full counting statistics (FCS)~\cite{schaller2014open}
of the departure process of the queue. We denote by the random variable $Q$ the number of customers that have left the
queue during the time interval $\tau$ and introduce the probabilities $\mathfrak{p}(j)$ of observing $Q=j$. At this point
we recall some essential results of the theory of FCS, which relate the cumulants $C_k$ of the distribution $\mathfrak{p}(j)$
to the generator $\gen$~\cite{Derrida-PRL-1998,Derrida2004}. The cumulant generating function $S(\xi)=\log\av{\rme^{\xi Q}}$
is proportional to the eigenvalue of smallest magnitude $\omega_0(\xi)$ of the deformed generator $\gen(\xi)$, i.e., $\omega_0(\xi)$
corresponds to the unique zero eigenvalue $\omega_0$ in the limit $\xi\rightarrow 0$. The deformed generator is defined by $\gen(\xi) = \gen^{(0)} + \rme^\xi\gen^{(1)}$, where $\gen^{(1)}$ contains all off-diagonal elements corresponding to transitions that increase $Q$
by one, and $\gen^{(0)}$ contains the remaining elements. For the steady state, the cumulants $C_j$ of the number of departures $Q$ in
the time interval $\tau$ are given by~\cite{Derrida-PRL-1998}
\begin{equation}\label{eigen}
	C_k = \frac{\partial^k S(\xi)}{\partial\xi^k}\Bigg|_{\xi=0} = \tau\frac{\partial^k \omega_0(\xi)}{\partial\xi^k}\Bigg|_{\xi=0}\hspace{40pt} k\geq 1\,.
\end{equation}
In particular, we obtain the average $C_1=\av{Q}$ and the variance $C_2 = \av{Q^2} - \av{Q}^2$. Since all cumulants
$C_k$ increase linearly with $\tau$ it is convenient to introduce the reduced cumulants $c_k = C_k/\tau$, where $c_1$
is the flow of customers and $c_2$ are the fluctuations.

The result in Eq.~\eref{eigen} is of limited use if it is not possible to find analytical expressions for the eigenvalue
$\omega_0(\xi)$. We avoid this problem by using the characteristic polynomial approach to FCS~\cite{Bruderer-NJP-14}. An
alternative method is based on identifying the cumulants from a Rayleigh-Schrödinger perturbation expansion for
the generating function~\cite{baiesi2009computation,Flindt-PRB-2010}.
The characteristic polynomial of the deformed generator $\gen(\xi)$ is defined by $P_\xi(x) = \det[x{\bf I} -\gen(\xi)]$,
with ${\bf I}$ the identity matrix. Generally, $P_\xi(x)$ of degree $d=K+1$ can be written in coefficient form as 
\begin{equation}\label{carpol1}
    P_{\xi}(x) = x^{d} + a_{d-1}x^{d-1} + \sum_{j=0}^{d-2}a_j(\xi)x^j\,.
\end{equation}
The coefficients $a_j(\xi)$ are given by the sum over the principal minors of order $d-j$ of the deformed generator
$\gen(\xi)$ and depend on the variable $\xi$, except for the coefficient $a_{d-1}$.

Let us establish the direct relation between the cumulants $c_k$ and the characteristic polynomial $P_\xi(x)$. First
note that the equality $P_{\xi}[\omega_0(\xi)] = 0$ holds by definition of the characteristic polynomial. By repeatedly
taking the total derivative of this equality with respect to $\xi$ and evaluating it at $\xi = 0$ we can generate the
(infinite) set of equations
\begin{equation}\label{deriv}
    \frac{\rmd^j P_{\xi}[\omega_0(\xi)]}{\rmd \xi^j}\Bigg|_{\xi=0} = 0\hspace{40pt} j\geq 1\,.
\end{equation}
These equations directly relate the cumulants $c_k$ to the coefficients $a_j$ and their derivatives
$a_j^{(\ell)}\equiv\partial^{\ell}_\xi a_j(\xi)|_{\xi = 0}$, which allows us to express the cumulants
$c_k$ in terms of $a_j$ and $a_j^{(\ell)}$. We obtain for the first three cumulants
\begin{equation}\label{khoff}
\eqalign{
	c_1 &= -\frac{a_0^\prime}{a_1} \\
	c_2 &= -\frac{1}{a_1}(2c_1a_1^{\prime} + 2a_2c_1^2 + a_0^{\prime\prime}) \\
	c_3 &= -\frac{1}{a_1}(6c_1^2a_2^{\prime} + 3c_2a_1^{\prime} + 3c_1a_1^{\prime\prime} + 6a_3c_1^3 + 6a_2c_2c_1 + a_0^{\prime\prime\prime})\,
}
\end{equation}
with $a_j^\prime\equiv a_j^{(1)}$ and so forth. The advantage of the characteristic polynomial approach is that analytical
expressions for the coefficients $a_j^{(\ell)}$ and hence for the cumulants $c_k$ may be found regardless of the dimension
of the system. It follows directly from Eqs.~\eref{khoff} that the cumulants are rational functions of the system parameters.

\begin{figure}[t]
\centering
\raisebox{4.5cm}{{\small\bf(a)}}\includegraphics[width=160pt]{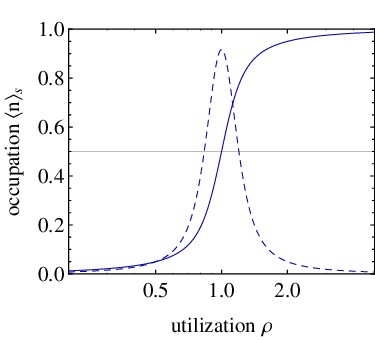}\hspace{20pt}
\raisebox{4.5cm}{{\small\bf(b)}}\includegraphics[width=165pt]{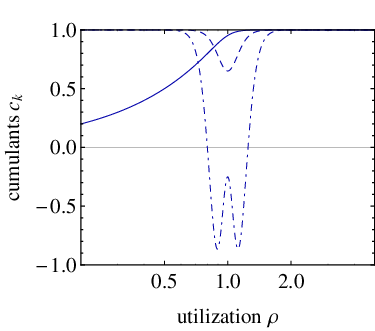}
\caption{The steady-state properties of $M/M/1/K$ queues: {\bf (a)}~The average occupation~$\av{n}$ (solid)
and variance $\av{\delta n^2}$ (dashed, $10\times$) as a function of the utilization $\rho$. The queue
exhibits a phase transition at $\rho=1$, characterized by a sudden change between the
low-occupation phase $\av{\delta n^2}\sim 0$ and high-occupation phase $\av{\delta n^2}\sim 1$.
Fluctuations of the occupation $\av{\delta n^2}$ are markedly increased at the transition $\rho=1$.
{\bf (b)}~The counting statistics of the departure process, parametrized by the cumulants, as a function of the
utilization $\rho$: $c_1/\mu$ (solid), $c_2/c_1$ (dashed) , $c_3/c_1$ (dash-dotted). The cumulants show strong
oscillations near the transition point $\rho=1$ , where the width of the oscillations decreases for increasing
$K$. The invariance of $\av{\delta n^2}$, $c_2/c_1$ and $c_3/c_1$ under the  transformation $\rho\rightarrow 1/\rho$ is emphasized
by the logarithmic scale on the x-axis. ($K=20$)}
\label{phase}
\end{figure}

\section{Properties of $M/M/1/K$ queues}

We illustrate the general approach by studying the behaviour of a standard queueing model, namely the $M/M/1/K$ queue
in Kendall's notation~\cite{Kleinrock-1976}. This model is comprised of a single server and a finite waiting space of size $K$ and the evolution
is fully specified by the constant arrival rates $\lambda_j\equiv\lambda$ and departure rates $\mu_j\equiv\mu$. A useful
quantity to describe the performance of the queue is the utilization $\rho\equiv\lambda/\mu$, which corresponds the load
of the server.

\subsection{Non-equilibrium steady-state occupations and phase diagram}

The steady-state probabilities for the $M/M/1/K$ queue are found from Eqs.~\eref{gensol} and given by
the geometric distribution~\cite{Kleinrock-1976}
\begin{equation}
	p_{\rm s}(j) = p_{\rm s}(0)\rho^j\hspace{40pt} j = 1,\ldots,K
\end{equation}
with
\begin{equation}\label{specsol}
  p_{\rm s}(0) = \cases{\frac{1-\rho}{1-\rho^{K+1}} & $\rho \neq 1$ \cr
							\frac{1}{1+K} & $\rho = 1$ \cr}\,.
\end{equation}
From the probabilities $p_{\rm s}(j)$ we find closed-form expressions for the average occupation
and variance
\begin{equation}
\eqalign{
	&\av{n} = \frac{\rho[1 - (K+1)\rho^K + K\rho^{K+1}]}{K(1-\rho)(1-\rho^{K+1})} \\
	&\av{\delta n^2} = \frac{\rho - (K+1)^2(1+\rho^2)\rho^{K+1} + 2K(K + 2)\rho^{K+2} + \rho^{2K+3}}{K^2(1-\rho)^2(1-\rho^{K+1})^2}\,,
}
\end{equation}
where both expressions are valid for $\rho\neq 1$. At the critical load $\rho = 1$ the two quantities take
the values $\av{n} = \frac{1}{2}$ and $\av{\delta n^2} = \frac{1}{12}(1 + \frac{2}{K})$.

Figure~\ref{phase} shows the dependence of the average occupation $\av{n}$ and the fluctuations $\av{\delta n^2}$
on the utilization $\rho$. We see that the system undergoes a non-equilibrium phase transition at $\rho=1$,
which is characterized by a sudden change between a low-occupation phase $\av{n}\sim 0$ and a high-occupation
phase $\av{n}\sim1$. The transition is smooth for finite values of $K$ and becomes sharp in the limit
$K\rightarrow\infty$. The corresponding phase diagram in the $\mu$-$\lambda$~plane is
therefore divided into two complementary regions separated by the phase boundary $\lambda = \mu$ (cf.~Fig.~\ref{scheme}).
Moreover, we observe that the occupation $\av{n}$ exhibits large fluctuations $\av{\delta n^2}$ in the vicinity
of the transition point, which is an additional indicator of the phase transition.

We point out that the $M/M/1/K$ queueing model is an effective description of the TASEP for particles in two cases: First,
if the hopping rate of the particles is significantly larger than the feeding and exit rates~\cite{Kolomeisky-JPA-1998}.
Second, if there is a slow bond that partially blocks hopping of particles~\cite{janowsky1992finite,romano2009queueing}.
The particles queue up in both cases and the dynamics of the TASEP is effectively described by the domain wall between
the full and empty region.

\subsection{Cumulants of the departure process}\label{cumulants}

Complementary to the analysis of the occupation of the queue we now focus on the flow
of customers, specifically on the statistics of the departure process of customers.
The analysis of the counting statistics of the departure process of the $M/M/1/K$
queue starts with the deformed generator
\begin{equation}\label{deformed}
\gen(\xi) = \left(
  \begin{array}{ccccc}
    -\lambda & \mu\rme^\xi &  &  & 0 \\
    \lambda & -(\lambda + \mu) & \mu\rme^\xi  &  & \\
     & \ddots & \ddots & \ddots & \\
     &  & \lambda & -(\lambda + \mu) & \mu\rme^\xi \\
     0 &  &  & \lambda & -\mu
  \end{array}
\right)\,.
\end{equation}
The complete characteristic polynomial $P_\xi(x)$ of degree $d$ contains all the information about the
counting statistics. However, we present explicit results only for the first three cumulants for which
part of the coefficients $a_j^{(\ell)}$ are needed (listed in~\ref{app}). Using Eqs.~\eref{khoff} we
then obtain the closed-form expressions for the first three cumulants ($\rho\neq 1$)
\setlength{\medmuskip}{3mu}
\begin{equation}\fl\label{excum}
\eqalign{
	c_1 &= \lambda\,\frac{1 - \rho^{K}}{1 - \rho^{K+1}} \\
	c_2 &= \lambda\,\frac{1 + \rho^{K+1}}{(1 - \rho^{K+1})^3}\left[1 - (2K+1)(\rho^{K} - \rho^{K+1}) - \rho^{2K+1}\right] \\
	c_3 &= \frac{\lambda}{(1 - \rho)(1 - \rho^{K+1})^5}\left[(1 - \rho)(1 - \rho^{5K+4}) + u_1(\rho^{K} + \rho^{4K+5}) + u_2(\rho^{2K+1} +\rho^{3K+4})\right. \\
			&\hspace{15pt} + u_3(\rho^{3K+2} + \rho^{2K+3}) + u_4(\rho^{4K+3} + \rho^{K+2}) + u_5(\rho^{K+1} + \rho^{4K+4}) + u_6(\rho^{2K+2} +\rho^{3K+3})\!\left.\right].
}
\end{equation}
\setlength{\medmuskip}{4mu}
The coefficients $u_j$ are quadratic polynomials in $K$, namely
\renewcommand\arraystretch{1.25}
\begin{equation}
\begin{fl}
\begin{array}{lll}
 u_1 = -1 - 3K - 3K^2 & u_2 = -8 - 27K - 21K^2 &  u_3 = -24 - 51K - 21K^2 \\
 u_4 = -14 - 15K - 3K^2  & u_5 = 9 + 18K + 6K^2 & u_6 = 38 + 78K + 42K^2 \,.
\end{array}
\end{fl}
\end{equation}
For the cumulants exactly at the phase boundary $\rho=1$ we find the simplified expressions
\begin{equation}
\eqalign{
 c_1 &= \lambda\frac{K}{K+1} \\
 c_2 &= \lambda\frac{K(2K+1)}{3(K+1)^2} \\
 c_3 &= \lambda\frac{K(6 + 14K + 11K^2 -K^3)}{30(K+1)^3}\,.
}
\end{equation}
We note that similar expressions for cumulants of higher order in terms of the waiting space $K$ and the
utilization $\rho$ can in principle be obtained, however, they become increasingly cumbersome. Closed-form
expressions for the variance $c_2$ have been presented in Ref.~\cite{nazarathy2008asymptotic}, where the
drop of $c_2$ at the critical load $\rho = 1$ was referred to as BRAVO effect, which occurs in several queueing
models~\cite{nazarathy2011variance,al2011asymptotic}.

The dependence of the flow $c_1$ and higher-order cumulants on the utilization $\rho$ are shown in Fig.~\ref{phase}.
As expected, the flow $c_1$ increases with $\rho$ and saturates at the value $\mu$ in the regime $\rho\gg 1$. The
behaviour of the higher-order cumulants is more remarkable: We see that all cumulants (normalized by the flow $c_1$)
show pronounced oscillations at the non-equilibrium phase transition, whereas the departure statistics is approximately
Poissonian away from the phase transition. The fluctuations $c_2$ of the departure process are strongly suppressed
(sub-Poissonian) at the critical load $\rho = 1$ with $c_2/c_1=2/3$ in the limit $K\rightarrow\infty$. The skewness
$c_3$ becomes arbitrarily small for increasing $K$ so that the departure statistics $\mathfrak{p}(j)$ is strongly
skewed to the left. In summary, the different non-equibrium phases of the $M/M/1/K$ queue are reflected not only
in the occupation $\av{n}$, but also in the full counting statistics of the flow of customers through the system.

\begin{figure}[t]
\centering
\raisebox{4.5cm}{{\small\bf(a)}}\includegraphics[width=160pt]{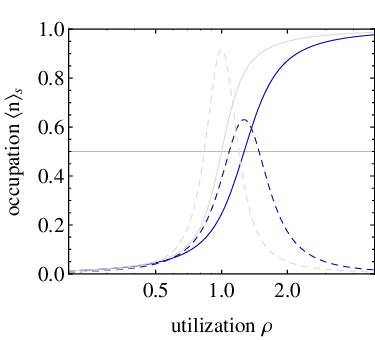}\hspace{20pt}
\raisebox{4.5cm}{{\small\bf(b)}}\includegraphics[width=165pt]{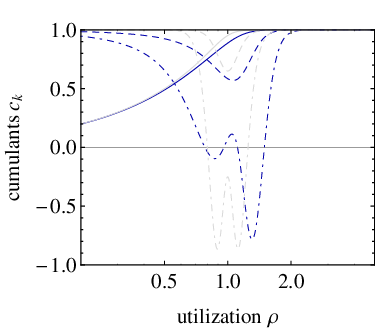}\vspace{10pt}\\
\raisebox{4.5cm}{{\small\bf(c)}}\includegraphics[width=160pt]{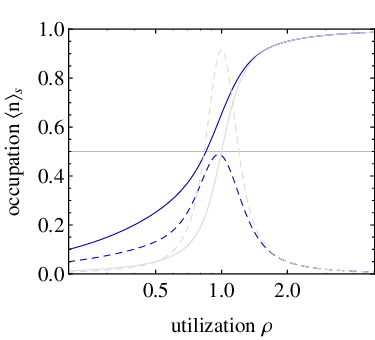}\hspace{20pt}
\raisebox{4.5cm}{{\small\bf(d)}}\includegraphics[width=165pt]{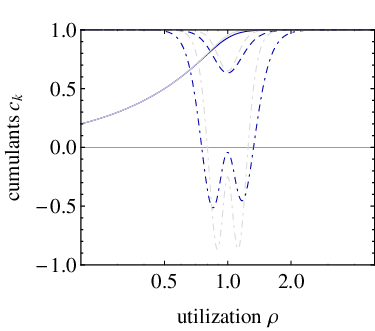}
\caption{The effect of discouragement and multiple servers on occupations and departure statistics: 
Discouragement of arrivals results {\bf (a)} in a shifted transition between low- and high-occupation phases and
{\bf (b)}~in a pronounced asymmetry of the skewness $c_3/c_1$.
The effect of multiple servers is mainly to broaden {\bf (c)}~the transition of the occupation and {\bf (d)} the
oscillations of the cumulants. Both modifications break the reciprocal symmetry, but preserve the general
features of the phase transition exhibited by the $M/M/1/K$ queue. (logarithmic scale on the x-axis, $K=20$, $m=10$,
$\theta=0.5$ and key as in Fig.~\ref{phase})}
\label{modifications}
\end{figure}

\subsection{Reciprocal symmetry}

The previous analysis of the $M/M/1/K$ queue suggests that the model possesses a reciprocal symmetry.
The underlying cause is that the queue of customers with arrival and service rates $\lambda$ and $\mu$ can be
interpreted as a queue of vacancies with the roles of $\lambda$ and $\mu$ reversed. One consequence of this
symmetry is that occupations and cumulants exhibit symmetries under the reciprocal transformation
$\rho\rightarrow 1/\rho$. Specifically, we find that the relevant quantities transform according to
$\av{n} \longrightarrow 1-\av{n}$, $\av{\delta n^2} \longrightarrow \av{\delta n^2}$ and $c_i/c_j \longrightarrow c_i/c_j$
for any two cumulants of order $i$ and $j$. In the next section we will break this reciprocal symmetry by
modifying the $M/M/1/K$ queue.

\section{Discouragement and multiple servers}

Considering the $M/M/1/K$ as a reference system we modify the properties of the queue to show that
the general statistical characteristics are robust with respect of the details of the queueing system.
The first modification are discouraged arrivals, i.e., the arrival rate is reduced if the filling of the
waiting space increases. We model this by generalizing the arrival rates to
\begin{equation}\label{disco}
	\lambda_j = \rme^{-\theta (j/K)}\lambda\,,
\end{equation}
where $j = 0,\ldots K$ are the number of customers in the queue and $\theta$ specifies the degree of discouragement.
We note that in general $\theta$ can be positive (discouragement) or negative (encouragement). In the present context
we assume $\theta > 0$, which may be physically interpreted as a repulsive force between particles in the system.
For $\theta = 0$ we obtain the standard $M/M/1/K$ queue to which the modified model is continuously connected.

A second modification is the introduction of multiple servers. For a multi-server queue with $m$ servers
and finite waiting space $K$, denoted by $M/M/m/K$, the departure rates are
\begin{equation}
	  \mu_j = \cases{ j\mu & $j=1,\ldots,m-1$ \cr
					m\mu & $j=m,\ldots,K$ \cr}\,.
\end{equation}
Note that the utilization for the $M/M/m/K$ queueing model is defined as $\rho\equiv\lambda/m\mu$.

We apply the methods described in Section~\ref{general} to the modified models to obtain
the occupation of the waiting space and the counting statistics of the departure process for specific
numerical parameters $\theta$, $m$ and $K$. Nevertheless, general analytical results, as for the $M/M/1/K$ queue,
can be obtained in principle. Figure~\ref{modifications} shows the average occupation of the waiting space and
the cumulants of the departure process for the modified models as a function of the utilization $\rho$.
Discouragement and multiple servers result in a broadened transition from the low-filling to the high-filling phase.
This broadening is also reflected in broadened oscillations of the lowest cumulants, in particular of dip of the
fluctuations $c_2$. Interestingly, the asymmetric shift of the local minima of the skewness $c_3$ allows us
to identify the single-server queue with discouragement.

\section{Conclusions}\label{conc}

The $M/M/1/K$ queueing model can be solved exactly: The occupation of the waiting space and the fluctuations
of the flow of departing customers have been explicitly calculated as a function of the utilization of the queue.
The model exhibits a load-dependent phase transition between a low-density and high-density phase which is
accompanied by oscillations of the cumulants of the departure process. These phenomena persist even if the
queueing model is generalized to include customer discouragement or multiple servers.

The variance and skewness of the departure statistics can in practice be obtained from repeated observations
of the number of customers Q that have left the queue during a time interval $\tau$~\cite{pawlikowski1990steady}.
In this way, phase transitions become observable solely based on monitoring departures and without knowing the
exact details of the states of the queue. Alternatively, one might want to know the size of the waiting space,
which can be inferred from the depth and width of the oscillations of the cumulants.

Let us consider the queueing model from a physics point of view, that is we identify customers with particles.
First, we find that the cumulants do not fall off as $1/K$ with the size of the waiting space $K$, in contrast
to one-dimensional models of length $L$, where a $1/L$-dependence is observed~\cite{bodineau2007cumulants}.
The queueing models studied here are therefore effectively zero-dimensional. Second, the currents exhibit
a dependence on the interaction between particles: The effect of the finite waiting space can be interpreted
as a repulsive short-range interaction, whereas discouraged arrivals correspond to a repulsive long-range
interaction. The skewness allows us to identify interactions, a feature that has been thoroughly studied
for electron transport~\cite{bagrets2006full}.

We expect that the oscillations of the cumulants in the vicinity of phase transitions are observable
in other queueing models. In addition, we conjecture that higher-order cumulants, which are rational
functions of the utilization, also exhibit oscillations. This is the natural generalization of the BRAVO
effect~\cite{nazarathy2008asymptotic,nazarathy2011variance,al2011asymptotic} to higher-order cumulants,
which may occur not only in queues but in other boundary-driven non-equilibrium systems.

\appendix

\section{Coefficients of the characteristic polynomial}\label{app}

\renewcommand\arraystretch{1.25}

The relevant coefficients of the characteristic polynomial $P_\xi(x)$ of degree $d = K + 1$ are
\begin{equation}\fl
\hspace{30pt}\begin{array}{ll}
a_1 = \mu^{d-1}\displaystyle\sum_{j=1}^d \rho^{j-1} & a_2 = \mu^{d-2}\displaystyle\sum_{j=1}^{d-1} \sym{j}{j}\,\rho^{j-1} \\
a_3 = \frac{1}{4}\mu^{d-3}\displaystyle\sum_{j=1}^{d-2}\sym{j}{j}\sym{j+1}{j+1}\,\rho^{j-1} & a_0^\prime = -\mu^d\displaystyle\sum_{j=1}^{d-1}\rho^j \\
a_1^\prime = -2\mu^{d-1}\displaystyle\sum_{j=1}^{d-2}\sym{j+1}{j}\,\rho^j & a_2^\prime = -\frac{3}{4}\mu ^{d-2} \displaystyle\sum_{j=1}^{d-3}\sym{j+1}{j}\sym{j+2}{j+1}\,\rho^j \\
a_0^{\prime\prime} = a_0^\prime + 2\mu^d\displaystyle\sum_{j=2}^{d-2}\sym{j+1}{j-1}\,\rho^j &
a_1^{\prime\prime} = a_1^\prime + \frac{3}{2}\mu ^{d-1}\displaystyle\sum_{j=2}^{d-3}\sym{j+1}{j-1}\sym{j+2}{j}\,\rho^j \\
a_0^{\prime\prime\prime} = 3a_0^{\prime\prime} -2a_0^\prime - \frac{3}{2}\mu^d\displaystyle\sum_{j=3}^{d-3}\sym{j+1}{j-1}\sym{j+2}{j-2}\,\rho^j 
\end{array}
\end{equation}
with the notation $\sym{j}{k}\equiv (d-j)k$.


\section*{References}

\bibliographystyle{with_titles}

\bibliography{arXiv_v1}


\end{document}